\documentclass{ws-ijmpb}

\usepackage{pifont}
\usepackage{graphics}

\begin{document}

\markboth{S. Haindl, M. Kidszun, F. Onken, A. Mietke, T. Thersleff}
{Lessons from Oxypnictide Thin Films}

\catchline{}{}{}{}{}
%
%%%%%%%%%%%%%%%%%%%%%%%%%%%%%%%%%%%%%%%%%%%%%%%%%%%%%%%%%%%%%%%%%%%%

\title{Lessons from Oxypnictide Thin Films}

\author{Silvia Haindl 
%\footnote{Typeset names in 8~pt Times Roman, uppercase. Use the footnote to indicate the present or permanent address of the author.}
}

\address{IFW Dresden, Institute for Solid State Physics, Helmholtzstr. 20\\
01069 Dresden, Germany \footnote{S.Haindl@ifw-dresden.de} $^{\dag}$ \\}

\author{Martin Kidszun}

\address{IFW Dresden, Institute for Solid State Physics, Helmholtzstr. 20\\
D-01069 Dresden, Germany \footnote{previous address: IFW Dresden, Institute of Metallic Materials, Helmholtzstr. 20
D-01069 Dresden, Germany}\\}

\author{Franziska Onken}

\address{TU Dresden, Faculty of Science, Department of Physics\\
D-01062 Dresden, Germany\\}

\author{Alexander Mietke}

\address{TU Dresden, Faculty of Science, Department of Physics\\
D-01062 Dresden, Germany\\}

\author{Thomas Thersleff}

\address{Uppsala University, Department of Engineering Sciences, Division of Applied Materials Science\\
Box 534 SE-751 21 Uppsala, Sweden $^{\dag}$ \\}

\maketitle

\begin{history}
pre-print version \\
appears in Int. J. Mod. Phys. B 27 (2013) 1230001
\end{history}

\begin{abstract}
First experiments on the growth of oxypnictide F-doped LaFeAsO thin films indicated an incomplete normal-to-superconducting transition and offered a work programme challenging to overcome possible difficulties in their fabrication. In this regard the possibility of an all \textit{in-situ} epitaxial growth appeared to be a matter of time and growth parameters. The following review clarifies that F-doped oxypnictide thin films are extremely difficult to grow by \textit{in-situ} PLD due to the formation of very stable impurity phases such as oxyfluorides (LaOF) and oxides (La$_2$O$_3$) and the loss of stoichiometry possibly due to incongruent evaporation of the target or re-evaporation of volatile elements at the substrate surface. However, the review also demonstrates that the employed two-step fabrication process for oxypnictide thin films has been successfully applied in the preparation of clean polycrystalline as well as of epitaxial thin films. Fundamental investigations on the upper critical field, its temperature dependence and its anisotropy contributed to an understanding of multiband superconductivity in oxypnictides.  
\end{abstract}


% - Introduction ------------------------------------------------
\section{Introduction}

Fe-based superconductors are a new class of high temperature superconductors containing Fe-pnictide and Fe-chalcogenide compounds with the common structural feature of tetrahedrally coordinated FeAs (or FeSe) layers that are supposed to be responsible for superconductivity. In the short history of Fe-based superconductors, a superconducting transition was first observed in the oxypnictides.\cite{Kam2006,Kam2008} At a first glance, the fact that an Fe containing compound shows superconducting properties is against intuition. This may also be one of the reasons why Jeitschko and co-workers, who synthesized a plethora of Fe-arsenides and Fe-phosphides already in the 1990s, were primarily interested in their magnetic properties.\cite{Zim1995,Que2000} 
Today, a huge number of different Fe-based superconductors are known and they are grouped into families of binary Fe-chalcogenides with composition \textit{1:1}, ternary Fe-pnictides and chalcogenides with composition \textit{1:1:1} or \textit{1:2:2}, and quaternary Fe-oxypnictides with composition \textit{1:1:1:1}. Also more complex oxypnictides\cite{Iva2010} show superconducting properties, however, this article will focus on the quaternary F-doped LnFeAsO oxypnictides (Ln = La, Sm).\cite{Iva2008} 

Several initial findings for the Fe-based superconductors have reached consensus: (1) The electronic band structure reveals several bands that cross the Fermi level resulting in a complex Fermi surface with different sheets, and it became clear that more than one electronic band is responsible for superconductivity.\cite{Sin2008} (2) The high critical temperatures up to 55 K in the oxypnictides\cite{Ren2008a} cannot be explained by a phonon-mediated electron-electron interaction.\cite{Boe2008} (3) A spin-density wave (SDW) state appears in most of the compounds and competes with the superconducting state suggesting that spin fluctuations are responsible for the superconducting pairing mechanism.\cite{dlC2008} Therefore, Fe-based superconductors can be classified as unconventional high-temperature superconductors of strong multiband nature. 

With a few exceptions, superconductivity appears upon charge carrier (electron or hole) doping, and the undoped compound is characterized as metallic or semi-metallic with a SDW developed. Superconductivity in the quaternary oxypncitides LnOFeAs (Ln = lanthanide) can be induced, for example, by F-doping\cite{Kam2008}, by oxygen deficiency\cite{Ren2008b}, by isovalent P-doping\cite{Wan2008}, by Co-doping\cite{Wan2009}, but also by external pressure\cite{Oka2008}. Today, the highest critical temperature in the oxypnictides is found in F-doped SmFeAsO that shows a $T_c$ of 55 K.\cite{Ren2008a} These high critical temperatures in the oxypnictides attracted enormous attention towards possible applications in which thin films do play a leading role.  

Despite the complexity of these compounds, first attempts of oxypnictide thin film fabrication started in 2008 using pulsed laser deposition (PLD).\cite{Hir2008,Bac2008} Immediately, the question about the F-content and, therefore, about the doping level turned out to be the first major experimental challenge. The first available LaFeAsO$_{1-x}$F$_x$ thin films neither showed any signs of superconductivity\cite{Hir2008} nor a $T_c$ above 20K and full resistive transitions.\cite{Bac2008} Until today, an all \textit{in-situ} deposition on heated substrates failed in growing high quality superconducting oxypnictide thin films. At present there is increasing activity in the oxypnictide thin film growth using molecular beam epitaxy (MBE).\cite{Kaw2009,Kaw2010,Ued2011a,Ued2012} Using MBE NdFeAsO$_{1-x}$F$_x$ and SmFeAsO$_{1-x}$F$_x$ thin films of good quality were obtained in the last years. There, F-doping was successfully achieved via diffusion from a cap layer either of NdOF, NdF$_3$ or SmF$_3$ respectively. We will not further discuss the peculiarities of MBE grown films here and want to refer to a summary in a recent article by Ueda \textit{et al.}\cite{Ued2011b}. 

This review is organized as follows: The first section is devoted to details of oxypnictide thin film preparation using PLD in combination with a post deposition annealing heat treatment (two-step process) resulting in polycrystalline and epitaxially grown LaFeAsO$_{1-x}$F$_{x}$ thin films. After a predominantly technical section \textit{2}, sections \textit{3} to \textit{5} review specific scientific problems that were investigated using LaFeAsO$_{1-x}$F$_{x}$ thin films. In section \textit{3} we will focus on the upper critical field of LaFeAsO$_{1-x}$F$_{x}$ thin films and its temperature dependence, whereas the anisotropy of the upper critical field is subject of section \textit{4} where the anisotropic Ginzburg-Landau scaling is used for its determination down to low temperatures. Section \textit{5} discusses potential applications for oxypnictide thin films. The review closes with a summary. 

% - Chapter 1 ---------------------------------------------------
\section{Pulsed Laser Deposition of Oxypnictide Thin Films}
  
\subsection{Principles of Pulsed Laser Deposition} 

Pulsed laser deposition (PLD) is a physical vapour deposition technique that uses a high power pulsed laser beam in order to vaporize material from a target surface.\cite{Chr1994,Eas2007,Che1988} Several important processes appear in PLD that can be described separately: (1) The interaction of the laser with the target, that is in principle a very complex process involving laser light absorption, target heating and plasma formation. (2) The vaporized material contains ions, electrons, and clusters - also called plasma plume, and it expands perpendicular to the target surface with velocities typically in the range of 10$^4$ ms$^{-1}$. Processes of plasma expansion under high vacuum or background gas atmosphere are, therefore, considered separately due to their longer duration compared to the duration of plasma formation. (3) Film nucleation and film growth on the substrate are considered as the final processes in PLD. 

% - Figure 1 ----------------------------------------------------------------------------
\begin{figure}[bt]
\centerline{\psfig{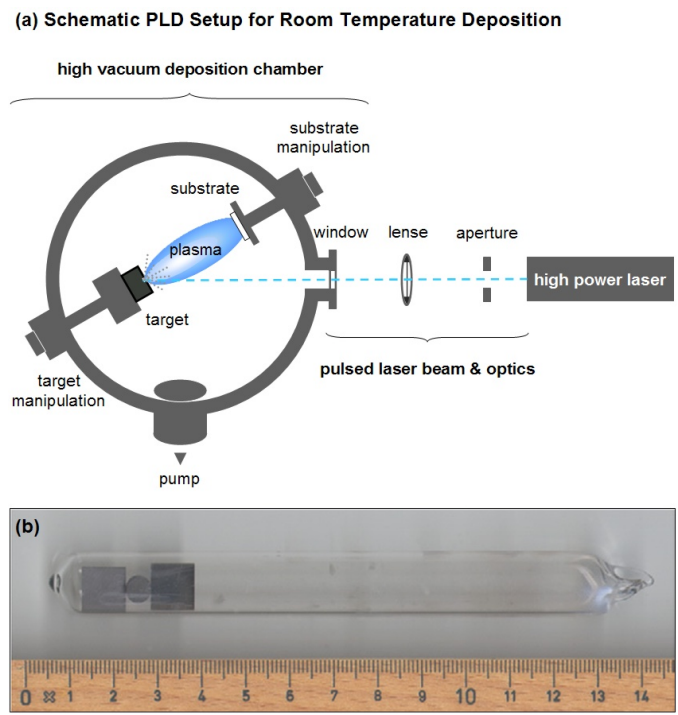}}
%\vspace*{8pt}
\caption{\textbf{(a)} Schematic setup for room temperature PLD consisting of a high power laser, the pulsed laser beam path and the deposition chamber. \textbf{(b)} (previously published as figure 2 in Kidszun \textit{et al.} \protect\cite{Kid2010a}) Evacuated quartz tube containing two as-grown LaFeAsO$_{1-x}$F$_{x}$ films and a sintered LaFeAsO$_{0.9}$F$_{0.1}$ pellet used during the \textit{ex-situ} heat treatment in the two-step fabrication process.}
\label{fig1}
\end{figure}
% ---------------------------------------------------------------------------------------

PLD as a thin film growth method owes its rise to the successful preparation of YBa$_2$Cu$_3$O$_{7-\delta}$ (YBCO) thin films.\cite{Dij1987} The main advantages of the PLD process are the following: (1) High deposition and growth rates. Using PLD thin films with thicknesses in the range of nanometers up to several hundreds of nanometers can be grown in a relatively short time of 1-100 minutes. (2) A stoichiometric transfer of the target material to the substrate that allows the growth of complex compounds from stoichiometric targets. (3) A highly directional material transfer from the target to the substrate. In vacuum atmosphere the plasma plume is strongly forward directed, i.e. normal to the target surface, even if the laser incidence is not normal. (4) The method is applicable in the growth of many different compounds such as nitrides, complex oxides including high-temperature superconductors, metals, polymer-metal compounds, and even fullerenes or other carbon-based materials. In addition, the pulsed process enables layer-by-layer growth and thus the synthesis of multilayer and quasi-multilayer films.\cite{Chr1994}  

The minimal PLD setup consists of a high power pulsed laser (KrF exciplex or Nd:YAG lasers with pulse durations between 5 and 50 ns), optics (lense, mirrors, window transparent to laser radiation) within the laser path and a deposition system that consists of a vacuum chamber with the possibility for the manipulation of subtrate and target holders (Figure~\ref{fig1}(a)). For a deposition at elevated temperatures substrate heating can be achieved by ceramic heaters directly placed above the substrate holder or by laser heating using diode or CO$_2$ lasers. During deposition a constant rotation of substrate and target holders is recommended in order to achieve better film homogeneity with respect to composition and thickness. 

\subsection{All In-Situ Depostion Not Yet Realized} 

Up to now it has proved difficult to grow superconducting (Ln)FeAsO$_{1-x}$F$_x$ (Ln = lanthanide) thin films by an all \textit{in-situ} PLD process. Therefore, all relevant parameters such as substrate temperature during deposition, repetition rate, laser fluence (or energy density at the target surface), target-substrate distance, vacuum conditions or the use of background gas for obtaining phase formation, texture, and optimal superconducting properties have not been investigated systematically so far. 

Besides the control of the F-content in the grown films, also the control of the As-content represents a major challenge. Stoichiometric material transfer is regarded as one of the advantages of the PLD process. However, the stoichiometric transfer is limited by a preferential ablation from the target, thermal evaporation at low energy due to different vapour pressures or segregation at the target surface as listed by Arnold and Aziz.\cite{Arn1999} Volatile elements like As are extremely sensitive to evaporation which can lead to further complications regarding the thin film stoichiometry.\cite{Schou2009} Off-stoichiometry then can be caused either by an already off-stoichiometric material transfer or by re-evaporation of the volatile elements from the substrate. In this case, the used target needs to be exposed to the laser irradiation a certain time until a steady-state in the ablation process is achieved or target compositions with excess of the volatile element will be necessary in order to compete with losses. Therefore, As-enriched targets have been employed by several groups for the growth of different Fe-based superconductors.\cite{Bac2008,Lee2009,Cho2009}  

\subsection{Two-Step Process} 

The successful fabrication process for oxypnictide thin films involves a two-step process: (1) room temperature PLD of a polycrystalline stoichiometric target followed by (2) an \textit{ex-situ} post deposition heat treatment in an evacuated quartz tube at temperatures of 940\,$^{\circ}\mathrm{C}$ to 960\,$^{\circ}\mathrm{C}$ (Figure~\ref{fig1}(b)).\cite{Bac2008,Kid2010a} Such a \textit{two-step} process including \textit{ex-situ} or \textit{in-situ} annealing is generally known, for example, from early fabrication of MgB$_2$ thin films, where the volatility of Mg\cite{Bri2001,Bla2001,Zha2001} posed problems, and also from the growth of Tl- or Hg-based cuprate thin films.\cite{Lee1988,Tsu1994} The two-step process has also been successfully applied in the growth of K-doped BaFe$_2$As$_2$ thin films.\cite{Lee2010K}

For the oxypnictide thin film growth the evacuated quartz tube contained additionally a pressed pellet of F-doped LaFeAsO (or SmFeAsO) of the same stoichiometry during the heat treatment. This sintered pellet turned out to be vital in order to keep the F-level and probably also the As-level within the thin film sample. 

Until today, an all \textit{in-situ} deposition on heated substrates failed in growing high quality superconducting oxypnictide thin films. However, a comparison with the few available oxypnictide thin film growth attempts using PLD demonstrated the superior success of employing the two-step process as first described by Backen \textit{et al.}\cite{Bac2008} (Table~\ref{depcond}). An all \textit{in-situ} deposition at elevated temperatures by Hiramatsu \textit{et al.}\cite{Hir2008} resulted in epitaxial but non-superconducting LaFeAsO thin films and thus emphasized the problem of F-losses. A two-step process including a post deposition heat treatment at temperatures between 900\,$^{\circ}\mathrm{C}$ and 1100\,$^{\circ}\mathrm{C}$ but without the additional use of pellets has not been successful in the growth of oxypnictide thin films, and therefore, this approach was abandoned by the same authors.\cite{Hir2008}

\subsubsection{Film Deposition at Room Temperature} 

The room-temperature deposition uses a standard PLD setup equipped with a KrF exciplex laser (with a wavelength of 248 nm and a typical energy density of 4 Jcm$^{-2}$ at the target surface). The setup is in principal equal to the one shown in Figure~\ref{fig1}(a), except for the motor for substrate manipulation that was demounted. The ablation process takes place in a high vacuum chamber (base pressure of 10$^{-6}$ mbar) with a target-substrate distance of 4 cm. LaFeAsO$_{0.9}$F$_{0.1}$ and LaFeAsO$_{0.75}$F$_{0.25}$ targets were synthesized based on the solid state reaction La$_2$O$_3$ + 2 LaO$_{0.5}$ + 4 FeAs $\rightarrow$ 4FeLaAsO, where the F content is controlled via the addition of LaF$_3$. The idea of using an overdoped target was to compete with F-losses during the PLD process as preliminary results indicated. 

Different substrates such as LaAlO$_3$, (La,Sr)(Al,Ta)O$_3$, and SrTiO$_3$ were used for oxypnictide thin film growth experiments, however, successful film growth was only achieved on LaAlO$_3$(001) substrates with a lattice parameter of $a$ = 3.82\,\AA. The lattice mismatch between the substrate lattice constant and the \textit{in-plane} lattice constant of LaFeAsO$_{0.9}$F$_{0.1}$ ($a$ = 4.028\,\AA) and of LaFeAsO$_{0.75}$F$_{0.25}$ (a = 4.018\,\AA) is about 5\% and a possible source for the interface layer formation of LaOF as described below in more detail. The lattice constants given here were determined from powder X-ray diffraction of the powders of the pellets that were used as target materials and in the \textit{ex-situ} heat treatment.

The Film thickness is primarily controlled by the number of pulses and the laser fluence (energy density at the target surface). Typically, film thicknesses varied between 100 and 1000 nm achieved by room temperature deposition. 

% - Table -------------------------------------------------------------------------------
\begin{table}[h!]   
\tbl{Deposition conditions for oxypnictide thin film growth by PLD}
{\resizebox{\linewidth}{!}{
\begin{tabular}{@{}lll@{}} \Hline \\[-1.8ex] 
reference & Hiramatsu \textit{et al.}\cite{Hir2008} & Backen \textit{et al.}\cite{Bac2008}\\[0.8ex] 
\hline \\[-1.8ex] 

\textit{laser} & {} & {} \\ 
source, wavelength & Nd:YAG (2$\omega$), 532 nm & KrF, 248 nm \\ 
repetition rate & 10 Hz & 10 Hz \\ 
energy density & 1.5 Jcm$^{-2}$ & 4 Jcm$^{-2}$ \\[0.8ex] 

\textit{target} & {} & {} \\ 
composition & LaFeAsO$_{0.9}$F$_{0.1}$ & LaFeAsO$_{0.75}$F$_{0.25}$ \\[0.8ex] 

\textit{substrates} \\ 
substrate temperature & 700 - 880\,$^{\circ}\mathrm{C}$ & room temperature \\
material, lattice parameter & MgO(100), 4.21\,\AA; & LaAlO$_3$(100), 3.82\,\AA \\ 
{} & (La,Sr)(Al,Ta)O$_3$, 3.87\,\AA & {} \\ 
{} & MgAl$_2$O$_4$, 8.09\,\AA & {} \\[0.8ex] 

all \textit{in-situ} results & heteroepitaxial growth & no phase formation \\
{} & but not superconducting & {} \\[0.8ex]

\hline \\[-1.8ex]

\textit{post heat treatment} & {} & {} \\[0.8ex]
 
\textit{1)} \textit{ex-situ} & {} & {} \\
temperature & 900 - 1100\,$^{\circ}\mathrm{C}$ & 1060\,$^{\circ}\mathrm{C}$ for 4h, \\
{} & {} & optimized to 950\,$^{\circ}\mathrm{C}$-960\,$^{\circ}\mathrm{C}$ \\
{} & {} & for 5-7h \cite{Kid2010a,Hai2010,Kid2010b} \\
use of a LaFeAsO$_{1-x}$F$_{x}$ pellet & \ding{55} & $\checkmark$ \\
results & not superconducting & polycrystalline, $T_c$ = 11 K \\
{} & {} & polycrystalline, $T_c$ = 28 K\cite{Hai2010} \\
{} & {} & epitaxial growth, $T_c$ = 25 K\cite{Kid2010b} \\[0.8ex]

\textit{2)} \textit{in-situ} & {} & {} \\
atmosphere, temperature & in 5 \% Ar/H at 400\,$^{\circ}\mathrm{C}$ for 0.5h & \ding{55}\\
{} & not superconducting & {}\\[0.8ex]

\Hline \\[-1.8ex] 
\end{tabular}}
}
\label{depcond}
\end{table}
% ---------------------------------------------------------------------------------------

\subsubsection{Ex-situ Heat Treatment} 

Using the above described two-step fabrication process the main process parameters concern the heat treatment rather than deposition conditions. These parameters are annealing time, maximum temperatures and temperature profile including heating and cooling ramps, pressure and control of the atmosphere in the sealed quartz tube, thermal stability of the substrate, etc. Thus, in several fabrication series the influence of these parameters was investigated and stepwise improved by primarily adjusting the temperature level and heating ramps as well as the duration of the heat treatment. Additional experiments were carried out in order to improve the pressure in the quartz tube and to investigate the role of the added LaFeAsO$_{0.9}$F$_{0.1}$ pellet. Good growth conditions were found for temperatures around 950\,$^{\circ}\mathrm{C}$ and an annealing time of 5-7 hours. However, individual experiments indicated that the optimal parameters are sensitive to the film thickness, especially when epitaxial growth of LaFeAsO$_{1-x}$F$_{x}$ should be achieved. 

An important point in the successful application of the two-step process was the use of an additional piece of a sintered pellet of same stoichiometry (Figure~\ref{fig1}(b)). The pellet prevents F and probably also As losses during the heat treatment of the as-grown films in the quartz tube. In addition, the reaction atmosphere inside the quartz tube may facilitate the phase formation process. High critical temperatures in the range between 25 K and 28 K were obtained when the heat treatment temperature was reduced from originally 1060\,$^{\circ}\mathrm{C}$ to around 950\,$^{\circ}\mathrm{C}$.\cite{Kid2010a,Hai2010,Kid2010b} Furthermore, the two-step process resulted finally in epitaxially grown LaFeAsO$_{1-x}$F$_{x}$ and SmFeAsO$_{1-x}$F$_{x}$ thin films.\cite{Kid2010b,Kid2011,Nai2011}     

In processes depending on a heat treatment the suitability of the substrate is not only given by its lattice misfit, but also determined by its chemical stability. Until today, detailed chemical reactions during the heat treatment are not resolved, but it is known that F$_2$ reacts with moisture and forms hydrogen fluoride, HF, that etches the quartz tube used in the \textit{ex-situ} heat treatment and/or the substrates. It is therefore regarded as necessary to avoid moisture and to evacuate the quartz tube before sealing. 
 
\subsection{Impurities, Film/Substrate Interface, and Doping} 

In all attempts of LaFeAsO$_{1-x}$F$_{x}$ thin film growth various impurity phases occured. Whereas LaAs formation was found in \textit{in-situ} grown films by Hiramatsu et al.\cite{Hir2008}, the \textit{in-situ} process in Backen et al.\cite{Bac2008} resulted mainly in La$_2$O$_3$ and LaOF impurity phases that seem to prevent any LaFeAsO$_{1-x}$F$_{x}$ phase formation.\cite{Kid2010a} 	

\begin{figure}[bt]
\centerline{\psfig{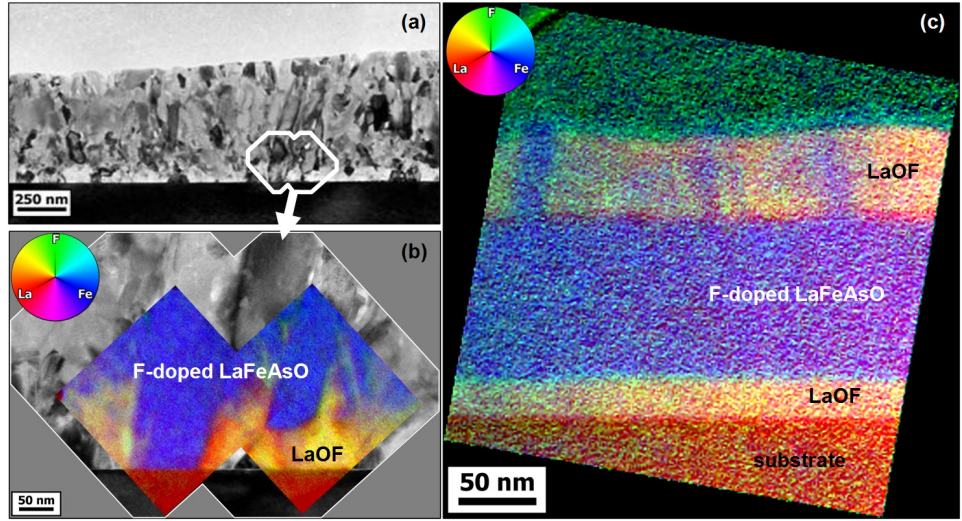}}
%\vspace*{8pt}
\caption{\textbf{(a)} Overview bright field TEM image of a polycrystalline LaFeAsO$_{1-x}$F$_x$ thin film. The white framed region is displayed in the figure below. \textbf{(b)} 3-window EFTEM (of core loss edges of F(K), La(M4,5), and Fe(L2,3) resulting in an RGB image) of the in (a) indicated region reveals LaOF grains at the film/substrate interface \textbf{(c)} (previously published as figure 5 in the supplement of Kidszun \textit{et al.} \protect\cite{Kid2011}) 3-window EFTEM (of core loss edges of F(K), La(M4,5), and Fe(L2,3) resulting in an RGB image) of the epitaxially grown LaFeAsO$_{1-x}$F$_x$ thin film reveals LaOF impurity phases at the film/substrate interface and on the film top. The lamellae were prepared by Focused Ion Beam (FIB) \textit{in-situ} lift-out technique and thinned to electron transparency using a 5 kV Ga$^{+}$ ion beam. TEM investigations were performed in a C$_S$-corrected FEI Titan 80-300 TEM operating at 300 kV at IFW Dresden.}
\label{fig2}
\end{figure}

The two-step fabrication process is also not free of impurity phase formation. There, the parameters of the \textit{ex-situ} heat treatment have a strong influence on the phase formation. Typical impurity phases like LaOF and La$_2$O$_3$ were identified in X-ray analysis carried out on the grown LaFeAsO$_{1-x}$F$_x$ films. Predominantly, LaOF competes with LaFeAsO$_{1-x}$F$_x$ phase formation as similarly seen in many attempts of synthesis and single crystal growth.\cite{Kar2009,McC2009} Analytical transmission electron microscopy (TEM), especially energy filtered TEM (EFTEM), resolves the main impurity, LaOF, at the film/substrate interface and on top of the sample (Figure~\ref{fig2}). These observations were made for the polycrystalline film (Figures~\ref{fig2}(a,b)) as well as for the epitaxially grown thin film (Figure~\ref{fig2}(c)). In the latter case also the LaOF layer at the film/substrate interface is grown epitaxially and, therefore, facilitates the epitaxial growth of LaFeAsO$_{1-x}$F$_x$. It can be assumed that in this case the LaOF layer helps to adjust the lattice misfit between the LaAlO$_3$ substrate and LaFeAsO$_{1-x}$F$_x$. 

The intermediate part of the thin films is dense and free of detectable impurities and thus allows performing electrical transport measurements with a defined current path. This microstructure is found to be typical for the two-step process used for film fabrication: the accumulated impurity concentration on top of the film can be explained by the growth process in the direction from the substrate to the top during the \textit{ex-situ} heat treatment. 

% - Figure 2 ----------------------------------------------------------------------------
% ---------------------------------------------------------------------------------------

A determination of the F-doping level within the grown films is difficult. For an indirect estimation of the F-content the superconducting transition temperature can be considered. $T_c$ values between 25 K and 28 K can be taken as an indicator for a F-content of approximately 10 at.\%. A direct measurement of the F-content in the thin films commonly involves destructive chemical analysis methods. A determination of the F-content by electron-energy-loss spectroscopy (EELS) on TEM specimens resulted in $x$ = 0.16, but this value seems to be overestimated because of overlapping signals from both, LaFeAsO$_{1-x}$F$_x$ and LaOF, phases. 

% - Chapter 2 ---------------------------------------------------
\section{Upper Critical Fields and Multiband Superconductivity}

\subsection{Multiband Superconductivity} 

The fact that more than one electronic band can contribute to superconductivity was originally and independently recognized by Suhl \textit{et al.}\cite{Suh1959} and Moskalenko\cite{Mos1959} in 1959 shortly after the development of the well known microscopic theory by Bardeen, Cooper and Schrieffer (BCS theory). Due to the overlap of different electronic bands at the Fermi level manifesting itself in different sheets of the Fermi surface, multiple gaps in the electronic excitation spectrum appear. A famous example of a two-band superconductor is MgB$_2$ with superconducting gaps on two separated Fermi sheets.\cite{Nag2001,Liu2001} 

Multiband superconductivity in the oxypnictides was considered theoretically soon after their discovery. Band structure calculations predict the Fermi surface consisting of four to five sheets, two electron cylinders around the M-point of the Brillouin zone, two hole cylinders and a hole pocket around the $\Gamma$-point arising from Fe 3d orbitals.\cite{Sin2008,Leb2007,Esc2009} Indeed, the majority of the experiments found at least two gaps different in size ($\Delta_{small}$ $\le$ 4 meV, $\Delta_{large}$ $\approx$ 7.5 meV).\cite{Evt2009,Gon2009} Point-contact spectroscopy was also carried out recently on LaFeAsO$_{1-x}$F$_x$ and La$_{1-x}$Sm$_x$FeAsO$_{1-x}$F$_x$ thin films grown by the two-step method, but only one clear gap around 6 meV was identified, a second gap feature remained unclear.\cite{Nai2011} Apart from spectroscopic probes, a manifestation of the multiband nature also appears in the temperature dependence of the upper critical field $H_{c2}(T)$, which can strongly deviate from the single band Werthamer-Helfand-Hohenberg (WHH) theory.\cite{Hel1964,Hel1966,Wer1966} It is therefore not surprising that one of the first claims of experimental evidence for multiband superconductivity in the oxypnictides was raised on the temperature dependence of $H_{c2}$ in LaFeAsO$_{1-x}$F$_x$.\cite{Hun2008} 

A decade earlier, Gurevich developed a handy theoretical description of $H_{c2}(T)$ applicable to the two-band superconductor MgB$_2$. His model is based on approximations for weak coupling (BCS limit) and information about the 2 $\times$ 2 coupling constant matrix $\Lambda$ with $\lambda_{11}$, $\lambda_{22}$ denoting the intraband and $\lambda_{12}$, $\lambda_{21}$ denoting the interband coupling constants.\cite{Gur2003,Bra2005,Bra2005e} For phonon mediated superconductors $\lambda_{ij}$ are precisely the electron-phonon coupling constants. In the Fe-based superconductors the interpretation may be generalized to electron-boson coupling constants because of the previously mentioned high critical temperatures\cite{Boe2008} and the possible magnetic interaction channel. 

\begin{figure}[t!]
\centerline{\psfig{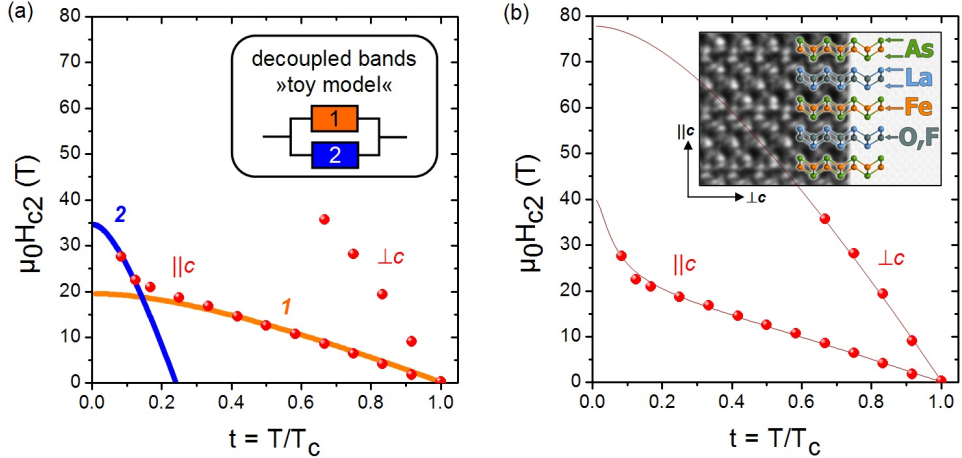}}
\vspace*{8pt}
\caption{\textbf{(a)} Measured temperature dependence of the upper critical field for an epitaxially grown LaFeAsO$_{1-x}$F$_x$ thin film (with $T_c$ = 25 K). Both major directions $H_{c2}\parallel c$ and $H_{c2}\perp c$ are shown. Qualitatively, a simple toy model can explain the measured temperature dependence of the upper critical field. A schematic description of two shunted electronic bands is given in the inset. \textbf{(b)} The same measured upper critical field values as in (a). Solid curves are specific solutions from Gurevich's two-band model. A HRTEM image resolves the layered structure of the LaFeAsO$_{1-x}$F$_x$ phase in the epitaxially grown film (inset, previously publisehd as inset in figure 1(b) in Kidszun \textit{et al.} \protect\cite{Kid2011}). Magnetotransport measurements in pulsed magnetic fields up to 42 T were carried out at IFW Dresden by A. Kauffmann and N. Kozlova in collaboration with J. Freudenberger. (Data previously published as figure 12 in the supplement of Kidszun \textit{et al.} \protect\cite{Kid2011})}
\label{fig3}
\end{figure}

The temperature dependence of $H_{c2}$ for an epitaxially grown LaFeAsO$_{1-x}$F$_x$ thin film was measured in pulsed magnetic fields up to 42 T for both directions, $H\parallel c$ and $H\perp c$ (as indicated by the dots in Figure~\ref{fig3}). The sudden upward curvature for $H_{c2}^{\parallel{c}}(T)$ at low temperatures ($T/T_c$ = $t <$ 0.2) can qualitatively be understood as the result of decoupled superconducting bands, where the weaker band is in the dirty limit having a decreased $T_c$ and an increased $H_{c2}$ due to impurity scattering. This so-called bilayer toy model of shunted electronic bands (first applied to MgB$_2$)\cite{Gur2007} was  qualitatively applied to the effective electron- and hole-band of the oxypnictide superconductor (lines in Figure~\ref{fig3}(a)). A similar trend for the temperature dependence of estimated $H_{c2}^{\parallel{c}}(T)$ was found by Hunte \textit{et al.} for $t <$ 0.5 in a polycrystalline LaFeAsO$_{0.89}$F$_{0.11}$ sample.\cite{Hun2008} 

% - Figure 3 ----------------------------------------------------------------------------
% ---------------------------------------------------------------------------------------

\subsection{Application of Gurevich's Two-Band Model to $H_{c2}(T)$} 

For the application of a simple two-band model, the following notation should be shortened by introducing w = det $\Lambda$, s = sgn(w), $\lambda_{\pm}$ = $\lambda_{11}$ $\pm$ $\lambda_{22}$ and $\lambda_{0}$ = ($\lambda_{-}^2$ + 4$\lambda_{12}$$\lambda_{21}$)$^{1/2}$. It can be deduced that a positive signum of w accounts for a stronger intraband coupling, whereas a negative signum of w is given for a stronger interband coupling. Neglecting interband scattering by nonmagnetic impurities $H_{c2}(t)$ is implicitely given by

\begin{equation} 
ln(t) = -\frac{1}{2}\left[U_1(h) + U_2(h) + \frac{\lambda_0}{w} \right] + s\sqrt{\frac{1}{4}\left( U_1(h) - U_2(h) - \frac{\lambda_-}{w} \right)^2 +\frac{\lambda_{12}\lambda_{21}}{w^2}} \label{equ1}
\end{equation}

\noindent with 

	\[U_{1,2}(h) = \Re\left(\psi\left[1/2 + (i + D_{1,2}/D_0)h\right]\right) - \psi(1/2)
\]

\noindent and 

\begin{equation}
H_{c2} = 2\Phi_0k_BT_cth/D_0\hbar, \label{equ2}
\end{equation}
\noindent where $h$ is the critical field parameter, $t = T/T_c$ is the reduced temperature, $D_0$ = $\hbar/2m$, and $\psi(x)$ is the digamma function (with $\Re$ denoting the real part). The complex argument in $U_{1,2}(h)$ accounts for paramagnetic effects. Neglecting them gives $U_{1,2}(h)$ = $\psi(1/2 + (D_{1,2}/D_0)h)$ - $\psi(1/2)$.\cite{Gur2003,Bra2005,Bra2005e} 

The diffusivities of the two bands $D_1$ and $D_2$ play an important role because according to their ratio, $\eta = D_2/D_1$, three cases can be distinguished: $\eta < 1$, $\eta = 1$ (resembling the one-band solution), and $\eta > 1$. For $\eta < 1$ the upper critical field shows a more or less pronounced upward curvature at low temperatures (as in our case above), whereas a positive curvature of $H_{c2}$ results from the case $\eta > 1$. Either knowledge or assumption of the electron-boson coupling constants, $\lambda_{ij}$, and the band diffusivities, $D_i$, (i, j indicating the band indices) is necessary to calculate $H_{c2}(t)$ with the help of equations (\ref{equ1}) and (\ref{equ2}). Here, impurity scattering is considered only by the band diffusivities and their ratio. More parameters or estimated values for an additional 2 $\times$ 2 impurity scattering matrix were omitted in the performed calculations.  

A least square fit procedure was implemented in a MATLAB\cite{Matlab} code which finds all possible solutions based on equation (\ref{equ1}) within a certain error range by varying the coupling constants. An initial step width of 0.01 for the coupling constants, $\lambda_{ij}$, in the range between 0.1 and 1 was used. To decrease the computing time the calculations were adjusted to minimize the deviation from the experimental data points. Assumptions on the diffusivities were made previous to the calculations for $H_{c2}^{\parallel{c}}$. The diffusivity ratio between the two bands was set initially to $\eta$ = 0.01 in order to adjust the sharp upward curvature of $H_{c2}^{\parallel{c}}$(T) at low temperatures. 

As demonstrated by the lines in Figure~\ref{fig3}(b) a simulation of the measured $H_{c2}(T)$ data points is possible using Gurevich's model,\cite{Gur2003}  however, the inversion of the problem, i.e. a quantification of the coupling constants determined from $H_{c2}(T)$ is impossible. There is no unique solution but rather a whole set of solutions. This set of solutions can however be characterized by $\lambda_{11}\cdot\lambda_{22} > \lambda_{21}\cdot\lambda_{12}$, i.e. the intraband coupling dominates over the interband coupling in an effective two-band model. 

Several points are noteworthy: 
(1) A drawback of the described procedure is the high number of parameters. Already in the two band scenario there are four coupling constants, $\lambda_{ij}$, and two diffusivity parameters, $D_i$, ($i,j$ = 1,2) with the impurity scattering matrix already neglected.
(2) A full description would of course involve a 4 $\times$ 4 or 5 $\times$ 5 coupling matrix and a much more complex theoretical approach yet not available. Here, we apply simply a 2 $\times$ 2 coupling matrix justified by the reduction to an effective electron band and an effective hole band in the oxypnictides. 
(3) Gurevich's model was applied also to $H_{c2}(T)$ of polycrystalline LaFeAsO$_{1-x}$F$_x$ where all coupling constants were set to $\lambda_{ij}$ = 0.5 ($\forall$ $i,j$) in order to reduce the number of unknown parameters.\cite{Koh2009} There, an upward curvature in $H_{c2}(T)$ was observed for low F-doping levels ($x$ = 0.05) and high F-doping levels ($x$ = 0.14), modelled with small diffusivity ratios $\eta$ = 0.1 (for $x$ = 0.05) and $\eta$ = 0.01 (for $x$ = 0.14).    
(4) There is also no information about the pairing symmetry, especially $s^{++}$ and $s^{\pm}$ cannot be distinguished primarily from $H_{c2}(T)$ since only the product $\lambda_{12}\cdot\lambda_{21}$ enters the equations and, therefore, cancels any sign of $\lambda_{12}$ ($\lambda_{21}$ respectively). Additional assumptions such as preferential interband coupling due to spin fluctuations for $s^{\pm}$ symmetry of the superconducting order parameter are neglected here, but have been applied in more complete descriptions involving more than two bands.\cite{Umm2009} Similar calculations using Gurevich's model were done to describe the temperature dependence of the upper critical field measured for single crystals of NdFeAsO$_{0.75}$F$_{0.25}$.\cite{Jar2008} Different coupling constant matrices (motivated by different pairing scenarios) were discussed there. A clear prediction however failed, because of the lack of $H_{c2}$ data in the low temperature and high field regions, where the scenarios could be distinguished. 

% - Chapter 3 ---------------------------------------------------
\section{Critical Current Scaling and $H_{c2}$ Anisotropy} 

\subsection{Anisotropic Ginzburg Landau Scaling Theory} 

Uniaxial electronic systems (superconductors with a layered structure) can be described by the effective electronic mass anisotropy, $\gamma_m = (m_{\parallel c}/m_{\perp c})^{1/2}$, where $m_{\parallel c}$ denotes the effective electron mass perpendicular to the layered structure and $m_{\perp c}$ denotes the effective electron mass parallel to the layers. It is well known that in the frame of the anisotropic Ginzburg-Landau theory the equation $\gamma_m = H_{c2}^{\perp c}/H_{c2}^{\parallel c} = H_{c1}^{\parallel c}/H_{c1}^{\perp c}$ ($\gamma_{Hc2} = \gamma_{Hc1}$) holds for an anisotropic single band superconductor.\cite{Tin1996} The angular dependence ($\theta$ denoting the angle between the applied magnetic field $H$ and the \textit{c}-axis of the uniaxial superconductor) of the upper critical field can thus be written as 

\begin{equation}
H_{c2}(\theta) = \frac{H_{c2}^{\parallel c}}{(cos^{2}(\theta)+\gamma_m^{-2} sin^{2}(\theta))^{1/2}}. \label{equ3}
\end{equation}

In the original work of Blatter, Geshkenbein, and Larkin\cite{Bla1992} it was shown that a scaling approach exists which maps a physical quantity, $Q$, obtained for an isotropic superconductor to the solution of an anisotropic superconductor by following a scaling rule that can be written as $Q$($\theta$,$H$,$T$,$\xi$,$\lambda$,$\gamma$,$\delta$) = $s_Q \cdot Q^{\star}$($\epsilon_{\theta}H$,$\gamma T$,$\xi$,$\lambda$,$\gamma\delta$). The following quantities are used: $\theta$ is defined above, $T$ is the temperature (strength of thermal fluctuations), $\xi$ and $\lambda$ are the coherence length and the penetration depth (values for the $ab$-plane), $\delta$ is the (scalar) disorder strength and $\epsilon_{\theta} = (cos^2(\theta)+\gamma_m^{-2} sin^2(\theta))^{1/2}$ with $\gamma_m^2$ the effective electronic mass anisotropy. For applied magnetic fields the scaling factor $s_Q$ is equal to 1/$\epsilon_{\theta}$, which leads to the above result for $Q \equiv H_{c2}(\theta)$ (with $Q^{\star} = H_{c2}^{\parallel c}$) in \ref{equ3}. 

Upon the assumption of \textit{uncorrelated disorder} (also called \textit{random pinning}), the scaling rule for the \textit{in-plane} critical current density is given by $Q \equiv J_c^{ip}(\theta ,H) = J_c^{H \parallel c}(\epsilon_{\theta} H)$ with $J_c^{H \parallel c} = J_c^{\star}$ $(\equiv Q^{\star})$ being the critical current density for the (equivalent) isotropic superconductor.\cite{Bla1994,Bla2003} This relation predicts the angular dependence of the \textit{in-plane} critical current density. Blatter and Geshkenbein write\cite{Bla2003}: \textit{Hence changing the direction of the magnetic field leads to a dependence of the \textit{in-plane} critical current density $J_c^{ip}(\theta, H)$ on the angle $\theta$ through the combination $\epsilon_{\theta}H$, resulting in sharp maxima of $J_c^{ip}(\theta, H)$ when the field is aligned with the superconducting planes and a sharpening of these maxima with increasing field amplitude H.} This scaling rule has been used in the discussion of the angular dependence of the \textit{in-plane} critical current densities by several authors, and it is valid in the weak collective pinning regime.\cite{Bra1993,Xu1994,Civ2004} The scaling method has also been used to extract uncorrelated pinning effects from correlated ones.\cite{Civ2004,Gut2007a,Gut2007b}  

The presence of \textit{correlated defects} (and thus \textit{correlated pinning}) might impose limitations on the application of the scaling procedure: The mass anisotropy for YBa$_2$Cu$_3$O$_{7-\delta}$ is known to be in the range of $\gamma_m$ = 5 to 7. Civale\cite{Civ2004} found a value of $\gamma_m$ = 5 and Guti\'{e}rrez \textit{et al.}\cite{Gut2007a} reported a value of $\gamma_m$ = 7 from the applied scaling theory on $J_c^{ip}(\theta,H)$ for YBCO thin films. However, as reported by the same authors\cite{Gut2007b} scaling was obtained in another YBCO thin film containing secondary phases of BaZrO$_3$ (BZO) only for a value of $\gamma_m$ = 1.5,  which was explained as \textit{decreased intrinsic anisotropy} for the BZO/YBCO composite. In the latter case, a discussion of \textit{pinning energy anisotropy} would be meaningful, because there, the \textit{in-plane} critical current density, $J_c^{ip}(\theta,H)$, is not only affected by the effective electronic mass anisotropy, but depends also strongly on extended pinning centers. 

% - Figure 4 ----------------------------------------------------------------------------
\begin{figure}[t!]
\centerline{\psfig{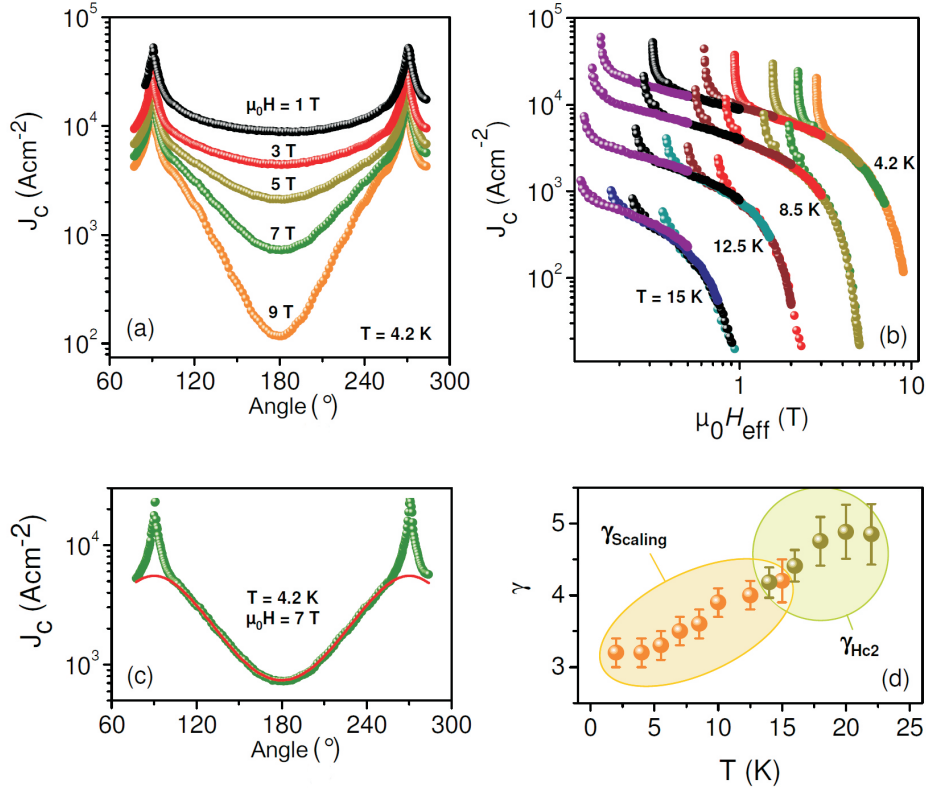}}
\vspace*{8pt}
\caption{\textbf{(a)} Angular dependence of the critical current densities at $T$ = 4.2 K for different applied magnetic fields (exemplarily shown for $\mu_0H$ = 1 T, 3 T, 5 T, 7 T and 9 T). $\theta$ = 180$^{\circ}$ corresponds to the configuration H$\parallel$c. \textbf{(b)} Scaling of $J_c$($\theta$) by using the scaling law for the effective magnetic field $H_{\text{eff}} = \epsilon_{\theta}H = H(cos^2(\theta) + \gamma_{\text{Scaling}}^{-2}sin^2(\theta))^{0.5}$. In the original theory $\gamma_{\text{Scaling}}^{2}$ corresponds to $\gamma_m^2 = m_{\parallel c}/m_{\perp c}$ the effective electronic mass anisotropy. Here, $\gamma_{\text{Scaling}}$ is evaluated for overlapping (i.e. \textit{scaling}) $J_c$ data. \textbf{(c)} Re-mapping exemplarily shown for $J_c$(4.2 K, 7 T). The red line of $J_c$($\theta$) indicates the $J_c$-anisotropy due to the mass anisotropy (if $\gamma_{\text{Scaling}}$ corresponds to $\gamma_m$) as it is obtained by the anisotropic Ginzburg-Landau scaling as shown in (b). Additional $J_c$ anisotropies for $\theta$ = 90$^{\circ}$ and $\theta$ = 270$^{\circ}$ that can not be described by the scaling theory originate from surface and intrinsic pinning effects. \textbf{(d)} Comparison between $\gamma_{\text{Scaling}}$ obtained from (b) and the measured $H_{c2}$-anisotropy and evidence for $\gamma_{\text{Scaling}}$ = $\gamma_{Hc2}$. Under the assumption of decoupled electron and hole bands we evaluate $\gamma_{m}$ = 4.8 for the electronic band dominating at high temperatures and $\gamma_{m}$=3.2 for the band dominating at low temperatures. $T_c$ of the film is 25 K. All transport critical current measurements were carried out in a commercial Physical Property Measurement System (PPMS) from Quantum Design in magnetic fields up to 9 T. (Data previously published as figure 2 in Kidszun \textit{et al.} \protect\cite{Kid2011})}
\label{fig4}
\end{figure}
% ---------------------------------------------------------------------------------------

% - Figure 5 ----------------------------------------------------------------------------
\begin{figure}[t!]
\centerline{\psfig{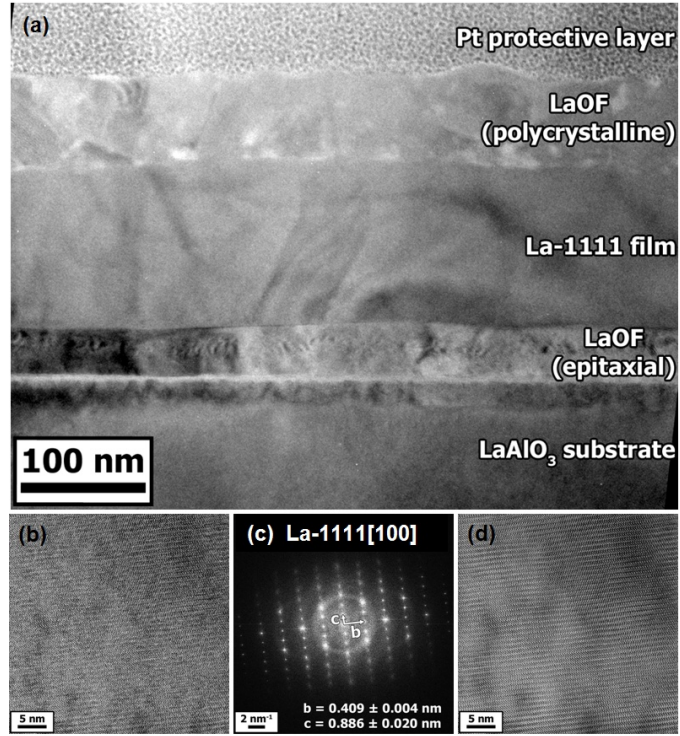}}
\vspace*{8pt}
\caption{\textbf{(a)} Bright field TEM overview image of the microstructure of the epitaxially grown LaFeAsO$_{1-x}$F$_x$ thin film (with the \textit{c}-axis oriented  \textit{out-of-plane}). The only extended secondary phases are LaOF layers at the film/substrate interface and on top of the oxypnictide thin film. Their influence to surface pinning effects can be seen for magnetic fields applied perpendicular to the \textit{c}-axis. \textbf{(b)} HRTEM micrograph of the LaFeAsO$_{1-x}$F$_x$ phase. \textbf{(c)} The Fast-Fourier-Transform (FFT) of image (b) indicates a highly crystalline film. The \textit{b}- and \textit{c}-axis can be measured and correspond with the values obtained by X-Ray diffraction. (d) Inverse FFT was filtered for spatial frequencies and superimposed with an opacity of 50\% on top of image (b). This serves as a noise reduction technique to emphasize the crystalline structure of the film. The investigated TEM-lamella was prepared by Focused Ion Beam (FIB) \textit{in-situ} lift-out technique and thinned to electron transparency using a 5 kV Ga$^{+}$ ion beam. All TEM investigations were performed in a C$_S$-corrected FEI Titan 80-300 TEM operating at 300 kV at IFW Dresden. (Previously unpublished.)}
\label{fig5}
\end{figure}
% ---------------------------------------------------------------------------------------

\subsection{In-Plane Critical Current Densities and $H_{c2}$ Anisotropy} 

It is well established, that the Fe-based superconductors are multiband superconductors and thus the relationship between $\gamma_m$ and $\gamma_{Hc2}$ (or the anisotropy of the coherence lengths, $\gamma_{\xi}$ and also the relationship between $\gamma_m$ and $\gamma_{Hc1}$ (or the anisotropy of the penetration depths, $\gamma_{\lambda}$, respectively) is generally unknown. As it was already shown for the two-band superconductor MgB$_2$, the anisotropies of upper and lower critical field are temperature dependent and, in general, do not match: $H_{c2}^{\perp c}/H_{c2}^{\parallel c} \neq H_{c1}^{\parallel c}/H_{c1}^{\perp c}$ ($\gamma_{\xi} \neq \gamma_{\lambda}$).\cite{Kog2003,Shi2003,Gur2004,Lya2004} 

The anisotropic Ginzburg-Landau scaling approach was applied to the epitaxially grown LaFeAsO$_{1-x}$F$_x$ thin film (Figure~\ref{fig4}).\cite{Kid2011} The measured angular dependence of the \textit{in-plane} critical current densities, $J_c^{ip}(H)$ is obtained upon rotation of the thin film in an external magnetic field where $\theta$ denotes the angle between the magnetic field direction and the \textit{c}-axis of the film (Figure~\ref{fig4}(a)). Note, that we do not probe the anisotropy of the critical current densities obtained from \textit{in-plane} transport and \textit{c}-axis transport measurements here.

Surprisingly, the scaling procedure works excellent in a wide temperature interval, despite the fact, that the oxypnictides are multiband superconductors (Figure~\ref{fig4}(b)). For reasons of clarity, we write the scaling factor as $\epsilon_{\theta} = (cos^2(\theta) + \gamma_{\text{Scaling}}^{-2}sin^2(\theta))^{0.5}$ and introduce technically $\gamma_{\text{Scaling}}$. In order to find the overlap of $J_c(\epsilon_{\theta}H)$ curves for a given temperature, the only parameter that has to be varied is the scaling parameter, $\gamma_{\text{Scaling}}$. Two questions arise here: Is the scaling parameter, $\gamma_{\text{Scaling}}$ governed by the $H_{c2}$ anisotropy ($\gamma_{Hc2}$, $\gamma_{\xi}$) or by the $H_{c1}$ anisotropy ($\gamma_{Hc1}$, $\gamma_{\lambda}$)? And, what is the relationship between the technically introduced $\gamma_{\text{Scaling}}$ and the mass anisotropy, $\gamma_m$? 

First, the scaling parameter $\gamma_{\text{Scaling}}$ obtained for different temperatures can be plotted as a function of temperature (Figure~\ref{fig4}(d)). Comparison between $\gamma_{\text{Scaling}}$ and the anisotropy of the upper critical field from pulsed magnetic field measurements (Figure~\ref{fig3}) shows an overlap at temperatures around 15 K. At present, further measurements are carried out in order to test the relationship between $\gamma_{\text{Scaling}}$ and $\gamma_{Hc2}$ in a broader temperature interval. Very recently van der Beek \textit{et al.}\cite{vdB2012} pointed out that two effects cause magnetic field and angular dependence of $J_c^{ip}(\theta,H)$. The first effect is the anisotropy of the pinning energy due to the anisotropy of the vortex line energy. The second effect is the anisotropy of the vortex core sizes (i.e. the anisotropy of the coherence lengths). In the case of core pinning of small and point-like defects (size comparable to the coherence length) the second effect dominates the angular dependence of $J_c^{ip}(\theta,H)$ leading to $\gamma_{Scaling}$ = $\gamma_{Hc2}$.  

Second, $\gamma_{\text{Scaling}}$ is interpreted as mass anisotropy $\gamma_m$ in the limits for $T \rightarrow T_c$ and $T \rightarrow 0$. A consistent explanation for this correspondence can be delivered, when the previous obtained results from the temperature dependence of $H_{c2}$ are considered. $H_{c2}^{\parallel c}(T)$ was discussed (in analogy to MgB$_2$) in the frame of decoupled effective electron and hole bands. The obtained $\gamma_m$ values in the limits of $T \rightarrow T_c$ (more precisely, $T \rightarrow T_{c,1}$, with $T_{c,1}$ denoting the critical temperature of the stronger band) and $T \rightarrow$ 0 (or $T \rightarrow T_{c,2}$, with $T_{c,2}$ denoting the critical temperature of the weaker band) would, therefore, correspond to effective electronic mass anisotropies of the dominating electronic bands. This result is also supported within a two-band Ginzburg-Landau theory, where in the discussed limits (neglecting impurity scattering) $\gamma_{m,1}$ = $\left(\gamma_{Hc2}\right)|_{Tc,1}$ and $\gamma_{m,2}$ = $\left(\gamma_{Hc2}\right)|_{Tc,2}$.\cite{Ask2006}

\begin{figure}[ht]
\centerline{\psfig{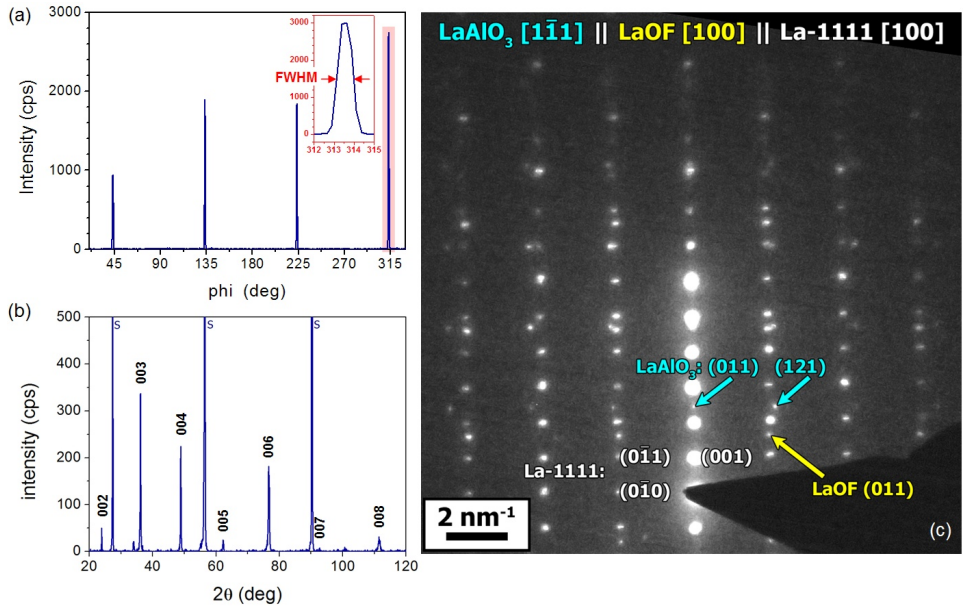}}
\vspace*{8pt}
\caption{\textbf{(a)} X-ray Phi-scan of the (112) reflection of the epitaxially grown LaFeAsO$_{1-x}$F$_x$ (La-1111) phase. The full-width-half-maximum (FWHM) is approximately 1$^{\circ}$. \textbf{(b)} X-ray diffraction pattern ($\theta-2\theta$ scan) of the epitaxially grown LaFeAsO$_{1-x}$F$_x$ thin film. $(00l)$ reflections of the LaFeAsO$_{1-x}$F$_x$ phase are indexed. Reflections from the LaAlO$_3$ substrate are denoted by \textit{S}. The X-ray analysis of the thin film was carried out using an X'Pert Philips X-ray Diffractometer with Cu K$\alpha$ radiation in (a) and Co K$\alpha$ radiation in (b). \textbf{(c)} A Selected Area Diffraction (SAD) of the oxypnictide film on the LaFeAsO$_{1-x}$F$_x$ phase and the LaOF phase indicates epitaxial growth. No further secondary phases are observed. The investigated TEM-lamella was prepared by Focused Ion Beam (FIB) \textit{in-situ} lift-out technique and thinned to electron transparency using a 5 kV Ga$^{+}$ ion beam. All TEM investigations were performed in a C$_S$-corrected FEI Titan 80-300 TEM operating at 300 kV at IFW Dresden. ((a) and (b) were previously published as figures 2(b) and (1) in Kidszun \textit{et al.} \protect\cite{Kid2010b}; (c) is previously unpublished.)}
\label{fig6}
\end{figure}

\begin{figure}[ht]
\centerline{\psfig{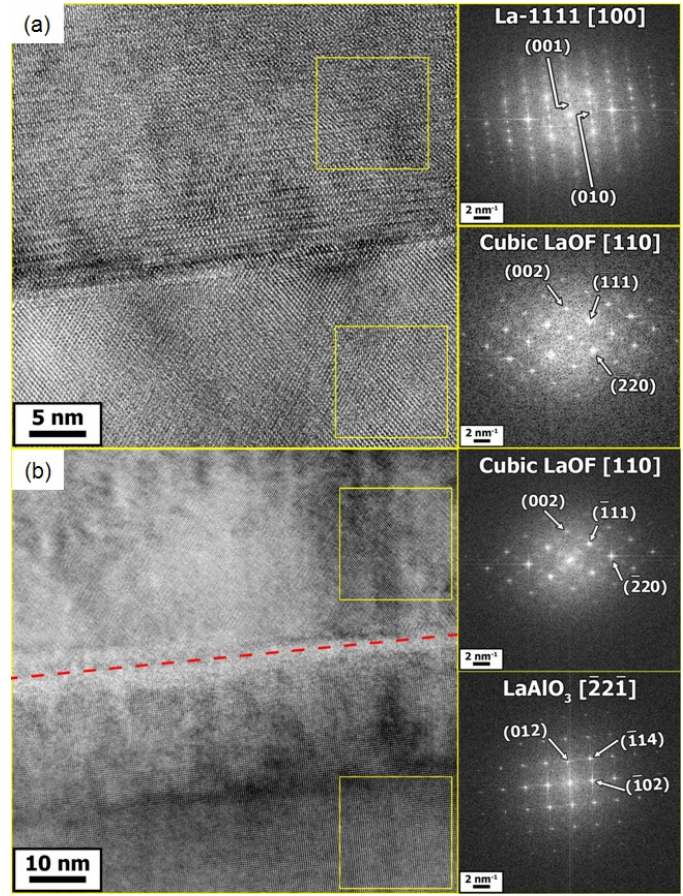}}
\vspace*{8pt}
\caption{\textbf{(a)} HRTEM at the epitaxial LaOF/LaFeAsO$_{1-x}$F$_x$ (La-1111) interface confirms the relationship La-1111[100] $\parallel$ LaOF[100] (FFT insets of the two regions). The interface is exceedingly clean with no secondary phases or reaction layers and is almost atomically sharp. \textbf{(b)} HRTEM at the LAO/LaOF interface reveals epitaxial growth of the LaOF phase (FFT insets of the two regions). A reaction layer appears within the LAO substrate.  This layer can be indexed as LAO albeit with broad spatial frequencies suggesting structural disorder within the LAO crystal lattice.  A likely explanation for this is the diffusion of F into the LAO substrate, which could substitute on the O site. The investigated TEM-lamella was prepared by Focused Ion Beam (FIB) \textit{in-situ} lift-out technique and thinned to electron transparency using a 5 kV Ga$^{+}$ ion beam. All TEM investigations were performed in a C$_S$-corrected FEI Titan 80-300 TEM operating at 300 kV at IFW Dresden. (Previously unpublished.)}
\label{fig7}
\end{figure}

The temperature dependence of the discussed anisotropies is commonly regarded to be a general feature of multiband superconductivity. The correspondence between $\gamma_m$ and $\gamma_{Hc2}$ can be understood qualitatively in terms of a decoupled bands scenario. The above investigated epitaxially grown LaFeAsO$_{1-x}$F$_x$ thin film with $T_c$ = 25 K shows a residual resistance ratio (RRR) of 6.8. It is important to note that the film is free of large extended defects (particularly, defects oriented parallel to the \textit{c}-axis): There is no indication for \textit{correlated defects} in the microstructure of the film (Figure~\ref{fig5}). Epitaxial growth is confirmed via X-ray analysis and TEM analysis (Figures~\ref{fig6} and ~\ref{fig7}). Again, the presence of only small point-like defects supports strongly the observation that the scaling parameter in the anisotropic Ginzburg-Landau scaling equals $\gamma_{Hc2}$.\cite{vdB2012} 

% - Figure 6 ----------------------------------------------------------------------------
% ---------------------------------------------------------------------------------------

% - Figure 7 ----------------------------------------------------------------------------
% ---------------------------------------------------------------------------------------

% - Chapter 4 ---------------------------------------------------
\section{Possible Applications For Oxypnictides}

The high transition temperatures of the oxypnictide superconductors triggered the question about their possible use for superconducting applications. Similar to the cuprates, the small coherence length in the oxypnictides can be regarded as responsible for a weak-link behaviour of the grain boundaries\cite{Hai2010}. Therefore, critical current densities in polycrystalline LaFeAsO$_{1-x}$F$_x$ films are about two orders of magnitude lower (approx. 10$^3$ Acm$^{-2}$ at 4.2 K) compared to critical current densities in epitaxially grown LaFeAsO$_{1-x}$F$_x$ films (approx. 7$\cdot$10$^4$ Acm$^{-2}$ at 4.2 K). Comparable values for critical current densities have been measured in LaFeAsO$_{1-x}$F$_x$ powder-in-tube (PIT) wires.\cite{Gao2008} SmFeAsO$_{1-x}$F$_x$ thin films and wires show slightly higher critical current densities compared to LaFeAsO$_{1-x}$F$_x$. 

The weak-link behaviour seems to be a general property for Fe-based superconductors, as it was reported also for Co-doped BaFe$_2$As$_2$ thin films grown on bicrystal substrates.\cite{Lee2009,Kat2011} Therefore, any possible high-current application as thin film conductor or PIT wire has to solve the technological problem of (biaxial) grain alignment.

In contrast to other Fe-based superconductors of composition \textit{1:1} or \textit{1:2:2}, the \textit{1:1:1:1} oxypnictides have an extended spacer oxide layer between the conducting FeAs layers. This gives rise to the possible formation of intrinsic Josephson junctions in these compounds, which can be applied in  emission sources for Terahertz radiation. In collaboration with Paul M\"{u}ller from University Erlangen-N\"{u}rnberg first experiments on current injection perpendicular to the FeAs layers in mesa structures of 5 $\times$ 5 $\mu$m$^2$ in lateral size for a 100 nm thin LaFeAsO$_{1-x}$F$_x$ film were carried out. The few published results are still controverse. On the one hand, Kashiwaya \textit{et al.}\cite{Kas2010} reported an intrinsic Josephson effect for PrFeAsO$_{0.7}$ single crystals. On the other hand, \textit{c}-axis transport measurements in oxygen deficient SmFeAsO$_{0.85}$ single crystals have not yet confirmed the presence of intrinsically Josephson coupled layers.\cite{Par2011}

\section{Summary}

Today, after five years of oxypnictide thin film growth, an all \textit{in-situ} PLD fabrication has still not been successful. The very few attempts so far resulted in the formation of impurity phases and stable oxides and oxyfluorides, but not in the growth of the so-called \textit{1:1:1:1} phase. Instead, a two-step process consisting of a conventional room temperature PLD and a subsequent \textit{ex-situ} post deposition heat treatment was successfully employed in the growth of LaFeAsO$_{1-x}$F$_x$ and SmFeAsO$_{1-x}$F$_x$ thin films. The main process parameters are however governed by the heat treatment rather than by the PLD itself. Therefore, important PLD parameters for an oxypnictide thin film growth have not yet been investigated in detail. Besides, the reproducibility of the oxypnictide thin film growth would be improved for an all \textit{in-situ} PLD process. The growth of oxypnictide thin films thus stays challenging.  

Polycrystalline and even epitaxial thin film growth was achieved by the two-step process. Still, the very stable LaOF impurity phase is present in the films. Microstructural investigations revealed that the LaOF phase is located generally at the film/substrate interface and on top of the film with the superconducting \textit{1:1:1:1} phase sandwiched in between. The overall clean and dense LaFeAsO$_{1-x}$F$_x$ layer can be perfectly analyzed by electrical transport measurements. 

As demonstrated for an epitaxially grown LaFeAsO$_{1-x}$F$_x$ thin film, the anisotropic Ginzburg-Landau scaling of the angular dependent critical current density can be applied. The resulting scaling parameter that corresponds to the square root of the effective electronic mass anisotropy in the limits for $T \rightarrow T_c$ and $T \rightarrow 0$ equals the upper critical field anisotropy for the respective temperatures. The temperature dependence of the effective electronic mass anisotropy is a general feature of multiband superconductivity. The multiband behaviour can also be probed in the temperature dependence of the upper critical field itself, which we were able to describe for LaFeAsO$_{1-x}$F$_x$ by a simplified two band model.    

Finally, the applicability of oxypnictides in superconducting devices faces technological limitations. Oxypnictides are not strongly two-dimensional and have effective electronic mass anistropies in the order of 5 to 7. On the one hand, electronic applications based on intrinsic Josephson junctions will be more difficult to realize than in Bi$_2$Sr$_2$CaCu$_2$O$_8$. On the other hand, applications based on high currents and pinning will still be affected by thermal fluctuations and flux flow at higher temperatures near $T_c$. In addition, the weak-link behaviour of grain boundaries is a disadvantage because grains have to be large and aligned in order to achieve a high critical current flow in power applications.

\section*{Acknowledgements}

The authors would like to thank M. Langer, J. Werner and G. Behr for target preparation, E. Reich, A. Reisner, K. Nenkov, U. Besold, and J. H\"{a}nisch for technical assistance. Furthermore, the authors would like to thank S. F\"{a}hler, G. Fuchs, and S.-L. Drechsler for fruitful discussions. Oxypnictide thin film growth and characterization was funded by the German Research Foundation (DFG) under project numbers HA5934/1-1 and HA5934/3-1.

\end{document}